\documentclass{ifacconf}
\usepackage{amsmath,amssymb}
\usepackage{subcaption}
\newcommand{\remarkend}{\hfill $\blacktriangle$} 
 \newtheorem{theorem}{Theorem}
  \newtheorem{proposition}{Proposition}
\newenvironment{proof}[1][Proof]{\textbf{#1.} }{\ \rule{0.5em}{0.5em}}
 
 \newtheorem{remark}{Remark}
\usepackage{graphicx}      
\usepackage{natbib}        
\begin{document}
\begin{frontmatter}

\title{Sliding Mode Control for a  Parabolic--Elliptic PDE System  with Boundary Perturbation} 

\author[First]{Moussa Labbadi} 
\author[Second]{Ilyasse Lamrani} 

\address[First]{Aix-Marseille University, LIS UMR CNRS 7020, 13013 Marseille, France (e-mail: \text{moussa.labbadi@lis-lab.fr})}
\address[Second]{ Department of Mathematics and Computer Science, École Marocaine des Sciences de l'ingénieur, EMSI, FEZ, Morocco.\\
TSI Laboratory, Department of Mathematics, Moulay Ismail
University, Faculty of Sciences, Meknes, Morocco
(e-mail: \text{lamrani.ilyasse.ts@gmail.com})}
\begin{abstract}                
In this paper, we address the robustness of parabolic--elliptic systems under boundary control. A sliding mode control strategy is proposed to reject matched perturbations. The stability analysis establishes finite-time convergence of the sliding manifold and exponential stability of the closed-loop system. Since the closed-loop system is discontinuous, we also prove its well-posedness. A numerical example is provided to validate the effectiveness of the proposed approach.
\end{abstract}
\end{frontmatter}
\section{Introduction}
Parabolic--elliptic systems naturally arise from the coupling of parabolic and elliptic partial differential equations (PDEs), where one component evolves dynamically while the other is assumed to reach a quasi-steady state. These systems have wide applications in physics, biology, and engineering~\citep{alalabi2024controller}, including lithium-ion and electrolytic cells \citep{ref1, ref2}, biological transport networks \citep{ref3}, and chemotaxis \citep{ref4, ref5}.

Mathematically, parabolic--elliptic systems strike a balance between realism and tractability: they allow rigorous analysis of existence, stability, blow-up, and numerical approximations, while the quasi-steady elliptic assumption simplifies computations. These systems can also be studied within the framework of differential-algebraic and Sobolev-type equations, enabling the use of semigroup theory to analyze well-posedness and stability. Seminal works have explored the construction of semigroups for differential-algebraic equations~\citep{Trostorff2018}, factorization methods for degenerate Cauchy problems~\citep{Thaller1996}, and linear Sobolev-type equations~\citep{Sviridyuk2012}. More recent studies address solvability of dissipative partial differential-algebraic equations~\citep{Jacob2022, morris2025boundary} and approaches using maximal monotone operators~\citep{fenza2025boundary}.

Stabilization via boundary control for coupled linear parabolic PDEs has been extensively investigated~\citep{baccoli2015boundary}. In~\citep{Coron2004}, the backstepping approach was used to stabilize linear coupled reaction--diffusion systems with constant coefficients, later extended to systems with variable coefficients~\citep{Krstic2008}. Koga et al.~\citep{Koga2010} studied boundary control of the one--phase Stefan problem, modeled by a diffusion equation coupled with an ODE, while feedback stabilization of PDE--ODE systems was analyzed in~\citep{Bastin2008}.

The backstepping method is notable for providing an explicit control law for PDEs without approximating the system, by mapping the original unstable system into an exponentially stable target system~\citep{fenza2025boundary}. This transformation redirects destabilizing terms to the boundary, where they are compensated by the control input. While coupling a stable parabolic equation with an elliptic equation might be expected to preserve stability, counterexamples show this is not always the case~\citep{AlalabiMorris}. Krstic and Smyshlyaev~\citep{Krstic2008} studied boundary stabilization of several coupled parabolic--elliptic systems, including linearized Kuramoto--Sivashinsky and Korteweg--de Vries equations, using two control inputs. Later work~\citep{AlalabiMorris, morris2025boundary} demonstrated stabilization of an unstable parabolic--elliptic system with Dirichlet boundary control in the presence of input delays, and addressed control and observer design via backstepping~\citep{AlalabiMorris, morris2025boundary}.

In the presence of a matched perturbation at the boundary, standard backstepping or Volterra-based control techniques cannot be directly applied due to the persistent influence of the disturbance on the boundary. This scenario requires more aggressive control strategies to ensure robustness. One suitable approach is Sliding Mode Control (SMC), which is well-known for its strong robustness properties and finite/fixed-time convergence. SMC has been extensively studied for infinite-dimensional systems and boundary control problems~\citep{orlov2002discontinuous, orlov2004robust, orlov1995sliding, orlov2013boundary, pisano2011tracking, pisano2012, pilloni2024sliding, utkin1992, polyakov2024finite, utkin2019control, balogoun2025sliding, balogoun2023, guo2012, guo2013, moreno2012, wang2015, chitour2024sliding, polyakov2025finite}.

Inspired by the aforementioned works, this paper proposes a sliding mode control strategy for parabolic–elliptic systems subject to matched perturbations. The main contributions are summarized as follows:
\begin{enumerate}
    \item We propose a sliding mode control strategy for parabolic–elliptic systems subject to matched perturbations.
    \item We establish finite-time stability of the sliding variable and exponential stability of the overall closed-loop system.
    \item Owing to the discontinuous nature of the resulting dynamics, well-posedness is nontrivial; therefore, we rigorously analyze the existence, uniqueness, and regularity of solutions.
   
\end{enumerate}
 A numerical example is provided to demonstrate the effectiveness of the proposed control scheme.
\section{Problem Statement}
Consider the coupled parabolic-elliptic system
\begin{subequations}
\begin{align}
u_t(x,t) &= u_{xx}(x,t) - \rho u(x,t) + \alpha v(x,t), \ x \in (0,1), \label{eq:u_eq2}\\
0 &= v_{xx}(x,t) - \gamma v(x,t) + \beta u(x,t), \ x \in (0,1), \label{eq:v_eq2}
\end{align}
\end{subequations}
with boundary conditions
\begin{subequations}
\begin{align}
u_x(0,t) &= 0, \qquad u_x(1,t) = \omega(t) + d(t), \label{eq:u_bc2}\\
v_x(0,t) &= 0, \qquad v_x(1,t) = 0, \label{eq:v_bc2}
\end{align}
\end{subequations}
and initial data
\begin{subequations}
\begin{align}
u(x,0) = u_0(x), \qquad v(x,0) = v_0(x). \label{eq:init2}
\end{align}
\end{subequations}
where $\omega(t)$ is the boundary control input, and $d(t) \in L_\infty[0,\infty)$ is an unknown but bounded disturbance acting at the boundary. 
In the absence of control input, the system can be reduced to a single evolution equation:
\begin{equation}
\begin{aligned}
u_t(x,t) &= u_{xx}(x,t) - \rho u(x,t)\\
& \qquad + \alpha \beta (\gamma I - \partial_{xx})^{-1} u(x,t),
\label{eq:reduced2}
\end{aligned}
\end{equation}
with Neumann boundary conditions $u_x(0,t)=0$, $u_x(1,t)=d(t)$.  

Define the operator
\begin{gather*}
A = \partial_{xx} - \rho I + \alpha \beta (\gamma I - \partial_{xx})^{-1}, \\
\quad D(A) = \{ u \in H^2(0,1) \mid u'(0)=u'(1)=0 \}.
\end{gather*}
\begin{theorem}[$\omega=d=0$]\citep{morris2023stabilization} \\
The eigenvalues of $A$ are
\begin{equation}
\lambda_n = -\rho + \frac{\alpha \beta}{\gamma + (n\pi)^2} - (n\pi)^2, \quad n=0,1,2,\dots,
\end{equation}
provided $\gamma \neq -(n\pi)^2$.
\end{theorem}
\begin{theorem}[Exponential Stability Criterion, $\omega=d=0$] \citep{morris2023stabilization}
The system is exponentially stable in $L^2(0,1)$ if
\begin{equation}
\rho > \frac{\alpha \beta}{\gamma},
\end{equation}
and the decay rate is bounded by the dominant eigenvalue $\lambda_0$.
\end{theorem}

For the nominal case $d=0$, the system can be mapped to a target exponentially stable system using a backstepping transformation of the form
\begin{equation}
\tilde u(x,t) = u(x,t) - \int_0^x k(x,y) u(y,t) \, dy,
\end{equation}
where the kernel $k(x,y)$ solves a suitable kernel PDE with Neumann boundary conditions.  
This ensures well-posedness and exponential stability of the closed-loop system \citep{morris2023stabilization}.

The goal is to design a boundary control law $\omega(t)$ such that the state $u(x,t)$ is stabilized in the presence of a bounded disturbance $d(t) \in L_\infty$, i.e.,
\begin{equation}
\|u(\cdot,t)\|_{L^2(0,1)} \to 0 \quad \text{as } t \to \infty,
\end{equation}
while maintaining robustness to the unknown boundary perturbation $d(t)$.  
This motivates the use of a sliding mode control strategy at the boundary to reject disturbances and enforce finite-time convergence.

\section{Main results}

\subsection{Robust Stabilization}\label{sec:SMC}
In this section, we design a boundary feedback of sliding-mode type to stabilize the coupled system. 
Unlike backstepping methods based on Volterra transformations, the SMC approach does not require constructing an invertible integral operator.

A possible strategy would be to rewrite system (\ref{eq:u_eq2})--(\ref{eq:v_bc2}) as a single equation in terms of $w(x,t)$, but this introduces a Fredholm operator
\[
\alpha\beta \int_0^1 g(x,y)\, w(y,t)\,dy,
\]
where $g(x,y)$ is the Green's function of $(\gamma I - \partial_{xx})^{-1}$, making a Volterra-type transformation impractical. 
Alternatively, a vector-valued transformation acting on both $w(x,t)$ and $v(x,t)$ could be used, but this becomes technically cumbersome. 
The sliding-mode boundary control avoids these difficulties and directly ensures robust stabilization.

We solve the elliptic equation \eqref{eq:v_bc2} under Neumann BCs to write:
\[
v(x,t) = \mathcal{K}[u(\cdot,t)],
\]
where \( \mathcal{K} : L^2(0,1) \rightarrow H^2(0,1) \) is the inverse of the elliptic operator with Neumann BCs:
\[
\mathcal{K} u = \left( \frac{d^2}{dx^2} - \gamma I \right)^{-1} (\beta u).
\]

Thus, the system reduces to:
\[
u_t(x,t) = u_{xx}(x,t) - \rho u(x,t) + \alpha \mathcal{K}[u(\cdot,t)].
\]

Inspired by~\citep{balogoun2025sliding}, let \( \psi \in H^1(0,1) \) be a fixed test function (for instance, smooth and compactly supported), and define the scalar sliding variable
\[
s(t) := \int_0^1 \psi(x) u(x,t) \, dx.
\]

We propose the boundary control input
\begin{align}\label{eq:u}
\omega(t) = -K\,\mathrm{sign}\big(s(t)\big)\;+\; \frac{\rho}{\phi(1)}\, s(t),
\end{align}
where \(K>0\) is a control gain to be chosen sufficiently large to dominate 
the unknown boundary disturbance \(d(t)\) and the remaining terms.

\begin{theorem}\label{thm:convergence_s}
Consider the parabolic subsystem and the sliding variable \(s(t)\) defined above.
Assume that:
    i) The disturbance is bounded: \( |d(t)| \le d_{\max} \);
   ii) The remainder term;
    \begin{gather*}
    R(t) := -\!\int_0^1 \psi_x(x) u_x(x,t)\, dx 
    \,+\, \alpha \!\int_0^1 \psi(x)\mathcal{K}[u](x,t)\, dx
     \end{gather*}
    is uniformly bounded: \( |R(t)| \le R_{\max} \).
   iii) \( \psi(1) \neq 0 \).

If the gain \(K\) satisfies
\begin{align}\label{eq:condition_K}
K > \frac{ |\psi(1)|d_{\max} + R_{\max} }{ |\psi(1)| },
\end{align}
then the sliding variable \(s(t)\) converges to zero in finite time; i.e.,
there exists a settling time \(T^* > 0\) such that
\[
s(t)=0 \qquad \text{for all } t \ge T^*.
\]
\end{theorem}
\begin{remark}
To relax the condition on $R(t)$, the boundary control input can be modified as
\begin{align}\label{eq:u1}
\omega(t) = -K\,\mathrm{sign}\big(s(t)\big) - \phi(1) R(t)+\frac{\rho}{\phi(1)} s(t),
\end{align}
In this context, the condition on the gain $K$ follows the classical sliding mode requirement:
\(
K > \frac{d_{\max}}{|\psi(1)|},
\)
this ensures the existence of the sliding manifold and the finite-time convergence of the system trajectories.
\remarkend
\end{remark}

\begin{proposition}\label{pro:exp}
Consider the parabolic subsystem
\[
u_t = u_{xx} - \rho u + \alpha \mathcal{K}[u], \quad x\in(0,1),
\]
with Neumann boundary conditions
\[
u_x(0,t) = 0, \qquad u_x(1,t) = \omega(t),
\]
and assume that the sliding manifold
\[
\mathcal{S} := \Big\{ u \in H^1(0,1) : s(t) = \int_0^1 \psi(x) u(x,t) dx = 0 \Big\}
\]
is reached in finite time. Suppose that \(\mathcal{K}: H^1(0,1)\to L^2(0,1)\) is bounded, i.e.,
\(\langle u, \mathcal{K}u \rangle \le C \|u\|^2\) for some \(C>0\), and \(\rho > \alpha C\). 

Then, the solution \(u(t)\) restricted to the sliding manifold converges exponentially to zero:
\[
\|u(t)\|_{H^1(0,1)} \le \|u(0)\|_{H^1(0,1)}\, e^{-\alpha_1 t},
\]
for some \(\alpha_1>0\).
\end{proposition}

\subsection{Well-posedness}
In this section, we investigate the well-posedness of the main system \eqref{eq:v_eq2}--\eqref{eq:v_bc2}, namely
\begin{align*}
	u_t(x,t)&= u_{xx}(x,t)-\rho\, u(x,t) + \alpha\, \mathcal{K}[u](x,t), x\in(0,1),\; t>0,\\[2mm]
	u_x(0,t) &= 0,\qquad  
	u_x(1,t) = \omega(t) + d(t), \qquad t>0,\\[2mm]
	u(x,0) &= u_0(x) \in L^2(0,1),
\end{align*}
where the boundary control $\omega$ is discontinuous~\eqref{eq:u}.
The nonlocal operator $\mathcal{K}$ is given by
$$
\mathcal{K}u = \left(\frac{d^2}{dx^2} - \gamma I\right)^{-1}(\beta u).
$$
Let $X_{-1}$ denote the completion of $L^2(0,1)$ with respect to the norm
$$
\|x\|_{X_{-1}} = \|(\sigma - A)^{-1} x\|_{L^2(0,1)},
$$
for some $\sigma \in \rho(A)$, the resolvent set of $A$.
Since $L^2(0,1)$ is reflexive, the space $X_{-1}$ can be identified with the dual space $(D(A^*))'$ endowed with the pivot space $L^2(0,1)$.

We define the unbounded boundary control operator associated with the Neumann boundary action at $x=1$ by
\begin{equation}
B : \mathbb{R} \to X_{-1}, 
\qquad 
Bu = \delta_1^{\prime} u,
\end{equation}
where $\delta_1^{\prime}$ denotes the derivative of the Dirac mass at $x=1$.
Since the boundary condition satisfies $u_x(1,t) = \omega(t) + d(t)$, we have
$$
Bu(t) = \omega(t) + d(t).
$$
Therefore, system \eqref{eq:v_eq2}--\eqref{eq:v_bc2} can be written in the following abstract form:
\begin{equation}
\begin{cases}
	\dot{u}(t) = A u(t) + B\big(\omega(t) + d(t)\big),\\[1mm]
	u(0) = u_0.
\end{cases}
\end{equation}
We now state the well-posedness result.
\begin{theorem} \label{thm:wellposed_disturbance}
	Suppose that
	
	i) $A$ generates an analytic $C_0$-semigroup $(S(t))_{t\ge 0}$ on $L^{2}(0,1)$;
		ii) $d \in L^{\infty}(0,\infty)$;
		ii) $B$ is admissible for the semigroup $(S(t))_{t\ge 0}$. 
	Then, for every $u_0 \in L^{2}(0,1)$, there exists a Filippov mild solution 
	\[
	u \in C([0,T]; L^{2}(0,1))
	\]
	to the differential inclusion
	\[
	\dot{u}(t) \in A u(t)+B(F[s(t)]+d(t)), \qquad u(0)=u_0,
	\]
	where $F[s]=-K\,\operatorname{sign}(s)$ and $s(t) = \langle \psi, u(t) \rangle$. Equivalently, there exists a measurable selection 
	$h(t)\in F[s(t)]$ for a.e.\ $t$ such that
	\[
	u(t)=S(t)u_0+\int_0^t S(t-s) B(h(s)+d(s))\, ds, \qquad t\in [0,T].
	\]
\end{theorem}

\section{Proofs of the main results}
\textbf{Proof of Theorem~\ref{thm:convergence_s}:}\\
\begin{proof}
Differentiating the sliding variable gives
\[
\dot{s}(t) = \int_0^1 \psi(x) u_t(x,t)\, dx.
\]

Using the PDE dynamics for \(u\) and integrating by parts yields
\begin{align*}
\dot{s}(t)
&= -\!\int_0^1 \psi_x(x) u_x(x,t)\, dx 
   + \psi(1) u_x(1,t) \\
   & \qquad - \rho\, s(t)
   + \alpha \!\int_0^1 \psi(x)\mathcal{K}[u](x,t)\, dx .
\end{align*}

Using the perturbed boundary condition \(u_x(1,t)=\omega(t)+d(t)\), we obtain
\[
\dot{s}(t)
= \psi(1)\big(\omega(t)+d(t)\big) + R(t),
\]
where the term \(R(t)\) is defined in the theorem.

Substituting the control law~\eqref{eq:u} gives
\[
\dot{s}(t)
= -K\psi(1)\,\mathrm{sign}(s(t))
  + \psi(1)\,d(t)
  + R(t).
\]

Taking absolute values and using the bounds on \(d(t)\) and \(R(t)\), we obtain
\[
\dot{s}(t)\,\mathrm{sign}(s(t))
\le -K|\psi(1)| + |\psi(1)|d_{\max} + R_{\max}.
\]

Define the strictly positive constant
\[
\eta := K|\psi(1)| - \big(|\psi(1)|d_{\max} + R_{\max}\big),
\]
which is positive under condition \eqref{eq:condition_K}. Then
\[
\dot{s}(t)\,\mathrm{sign}(s(t)) \le -\eta.
\]

Noting that
\[
\frac{d}{dt}|s(t)| = \dot{s}(t)\,\mathrm{sign}(s(t)),
\]
we obtain
\[
\frac{d}{dt}|s(t)| \le -\eta.
\]

Integrating yields
\[
|s(t)| \le |s(0)| - \eta t.
\]

Thus, the settling time
\[
T^* = \frac{|s(0)|}{\eta}
\]
is finite, and \(s(t)=0\) for all \(t \ge T^*\). Once in sliding mode, the inequality 
\(\dot{s}(t)\,\mathrm{sign}(s(t)) \le -\eta < 0\) ensures invariance of the manifold.
\end{proof}

\textbf{Proof of Proposition~\ref{pro:exp}:}\\
\begin{proof}
Consider the Lyapunov candidate
\[
V(t) = \frac{1}{2}\|u(t)\|_{L^2(0,1)}^2.
\]

Differentiating and substituting the PDE gives
\[
\dot{V}(t) = \int_0^1 u\, u_t \, dx
= \int_0^1 u \big(u_{xx} - \rho u + \alpha \mathcal{K}u \big) dx.
\]

Integrating by parts and using \(u_x(0,t)=0\) and \(u_x(1,t)=\omega(t)\) yields
\[
\int_0^1 u u_{xx} dx = - \int_0^1 |u_x|^2 dx + u(1,t) \omega(t),
\]
so that
\[
\dot{V}(t) = -\|u_x\|^2 - \rho \|u\|^2 + \alpha \langle u, \mathcal{K}u \rangle + u(1,t) \omega(t).
\]

Since the sliding manifold has been reached, \(|u(1,t)\omega(t)|\) is bounded by
\(|u(1,t)\omega(t)| \le \varepsilon \|u\|^2 + \delta \|u_x\|^2\)
for sufficiently small \(\varepsilon,\delta>0\). Also, \(\langle u, \mathcal{K}u\rangle \le C \|u\|^2\). Thus,
\[
\dot{V}(t) \le - (1-\delta) \|u_x\|^2 - (\rho - \alpha C - \varepsilon) \|u\|^2.
\]

Choosing \(\varepsilon, \delta\) sufficiently small, and noting \(\rho > \alpha C\), we define
\(\alpha_1 := \min\{1-\delta, \rho - \alpha C - \varepsilon\} > 0\), yielding
\[
\dot{V}(t) \le - \alpha_1 \|u\|_{H^1(0,1)}^2.
\]

Standard Lyapunov arguments then imply exponential convergence:
\[
\|u(t)\|_{H^1(0,1)} \le \|u(0)\|_{H^1(0,1)}\, e^{-\alpha_1 t}.
\]
\end{proof}

\textbf{Proof of Theorem~\ref{thm:wellposed_disturbance}:}\\
\begin{proof}
	We begin by verifying that the operator
	$$
	A := \partial_{xx}-\rho I+\alpha \beta(\gamma I-\partial_{xx})^{-1}
$$
	generates an analytic $C_0$-semigroup on $L^2(0,1)$. Indeed, $A$ is a bounded perturbation of the Neumann Laplacian, which generates a strongly continuous analytic semigroup. Hence $A$ generates an analytic $C_0$-semigroup $(S(t))_{t\ge 0}$ (see \cite[p.~81]{pazy2012semigroups}).

	\noindent\textit{Regularization of the feedback.}
	We apply a standard regularization argument (see \citep{aubin2008differential}). Let $(\theta_n)_{n\ge 1}$ be a sequence of Lipschitz-continuous functions
$$
	\theta_n : \mathbb R \to \mathbb R,\quad
	\|\theta_n\|_\infty \le 1,\quad
	\theta_n(r) \to \operatorname{sign}(r) \ \text{for } r\neq 0,
$$
	with Lipschitz constants $L_n$. Define the regularized feedback
$$
	\omega_n(t) := -K\theta_n\big(s_n(t)\big)+ \frac{\rho}{\phi(1)} s_n(t), \
	s_n(t) := \langle \psi, u_n(t) \rangle_{L^{2}(0,1)}.
$$
	
	\smallskip
	\noindent\textit{Existence for the regularized problem.}
	For fixed $n$, consider the mild formulation:
	\begin{equation}\label{eq:un_mild}
	\begin{aligned}
		u_n(t)
		= S(t)u_0
		+ \int_0^t S(t-s) B\big(-K\theta_n(\langle \psi, u_n(s)\rangle)\\ +\frac{\rho}{\phi(1)}\langle \psi, u_n(s)\rangle + d(s)\big)ds.
	\end{aligned}
	\end{equation}
	Define $\theta_n : Y \to Y$, with $Y:=C([0,T];L^{2}(0,1))$, by
		\begin{equation}\label{eq:un_mil1d}
	\begin{aligned}
	(\theta_n u)(t)
	:= S(t)u_0
	+ \int_0^t S(t-s) B\big(-K\theta_n(\langle \psi,u(s)\rangle)\\ + \frac{\rho}{\phi(1)}\langle \psi,u(s)\rangle+ d(s)\big)\,ds.
	\end{aligned}
	\end{equation}
	By admissibility of $B$, there exists $C_T>0$ such that for all $h\in L^2(0,T)$,
	$$
	\Big\|\int_0^\cdot S(\cdot - s) Bh(s)\,ds\Big\|_{Y}
	\le C_T \|h\|_{L^2(0,T)}.
$$
	For $u,v\in Y$, the Lipschitz property of $\theta_n$ gives
	$$
	\|\theta_n u - \theta_n v\|_Y
	\le C_T K L_n \|\psi\|_{L^2}\sqrt{T}\,\|u-v\|_Y.
$$
	Choosing $T>0$ such that
	$$
	\kappa_n := C_T K L_n \|\psi\|_{L^2}\sqrt{T} < 1,
$$
	the operator $\theta_n$ is a contraction. Thus \eqref{eq:un_mild} admits a unique mild solution
	$u_n \in C([0,T];L^2(0,1))$. Extending by continuation, we obtain existence on any finite interval.
	
	\noindent\textit{Uniform bounds.}
	By \eqref{eq:un_mild} and admissibility,
	\begin{gather*}
	\|u_n\|_{C([0,T];L^{2})}
	\le \|S(\cdot)u_0\|_Y
	+ C_T\|-K\theta_n(\langle\psi,u_n\rangle) \\+ \frac{\rho}{\phi(1)}\langle \psi,u_n\rangle+ d\|_{L^2(0,T)}.
	\end{gather*}
	Using $|\theta_n|\le 1$ and $d\in L^\infty(0,T)$ we obtain
$$
	\|u_n\|_{C([0,T];L^{2})}
	\le \|S(\cdot)u_0\|_{Y}
	+ C_T\big(K\sqrt{T}+ \|d\|_{L^2(0,T)}\big)
	=: M_T,
$$
	where $M_T$ is independent of $n$. Hence $s_n(t)=\langle \psi,u_n(t)\rangle$ is uniformly bounded, and
$$
	\xi_n(t):=-K\,\theta_n(s_n(t))
$$
	is uniformly bounded in $L^\infty(0,T)$.

	\noindent\textit{Passage to the limit.}
	The sequence $h_n := \xi_n + d$ is bounded in $L^\infty(0,T)$, hence weak-$^\ast$ relatively compact. Thus, up to a subsequence,
	$$
	h_n \overset{*}{\rightharpoonup} h
	\qquad\text{in } L^\infty(0,T).
	$$
	Set $\xi := h - d$.
	
	Define the solution operator
	$$
	(\mathcal T g)(t)
	:= \int_0^t S(t-s) B g(s)\,ds.
	$$
	Since $A$ generates an analytic semigroup, $\mathcal T$ is compact from $L^2(0,T)$ into $C([0,T];L^{2}(0,1))$. Because $h_n$ is bounded in $L^\infty(0,T)$, it converges strongly:
	$$
	u_n = S(\cdot)u_0 + \mathcal T h_n \longrightarrow
	u := S(\cdot)u_0 + \mathcal T h
	\quad\text{in } C([0,T];L^{2}(0,1)).
	$$
	Since $s_n(t)=\langle \psi, u_n(t)\rangle$ and $u_n\to u$ uniformly, we obtain
	$$
	s_n \to s:=\langle \psi,u\rangle
	\quad\text{uniformly on }[0,T].
	$$
	The graph of the maximal monotone map $F(r)=-K\,\mathrm{sign}(r)$ is sequentially closed under strong convergence of inputs and weak-$^\ast$ convergence of outputs. Thus, from
	$$
	s_n \to s \ \text{strongly}, \qquad \xi_n \rightharpoonup^\ast \xi,
	$$
	we deduce
$$
	\xi(t) \in F[s(t)]
	\qquad\text{for a.e.\ } t\in [0,T].
$$
	Therefore,
	$$
	u(t)=S(t)u_0+\int_0^t S(t-s)B\big(\xi(s)+d(s)\big)\,ds
$$
	is a Filippov mild solution of the differential inclusion, which concludes the proof.
\end{proof}





\section{Numerical simulations}
The numerical simulations illustrate the effectiveness of the proposed sliding mode boundary control for the coupled parabolic-elliptic system. 

\begin{figure}[ht!]
    \centering
    \begin{subfigure}[b]{0.47\textwidth}
        \centering
        \includegraphics[width=\textwidth]{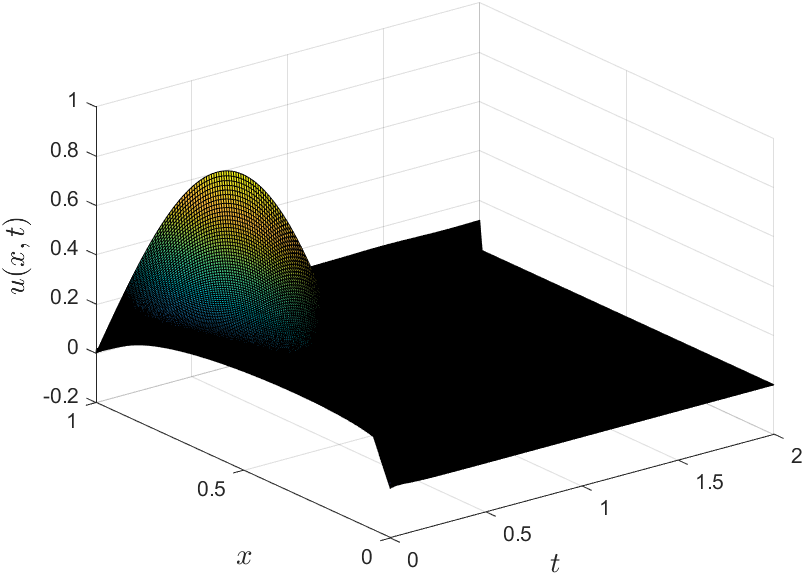}
        \caption{Trajectory of $u(x,t)$ before and after control.}
        \label{fig:u_traj}
    \end{subfigure}
    \hfill
    \begin{subfigure}[b]{0.47\textwidth}
        \centering
        \includegraphics[width=\textwidth]{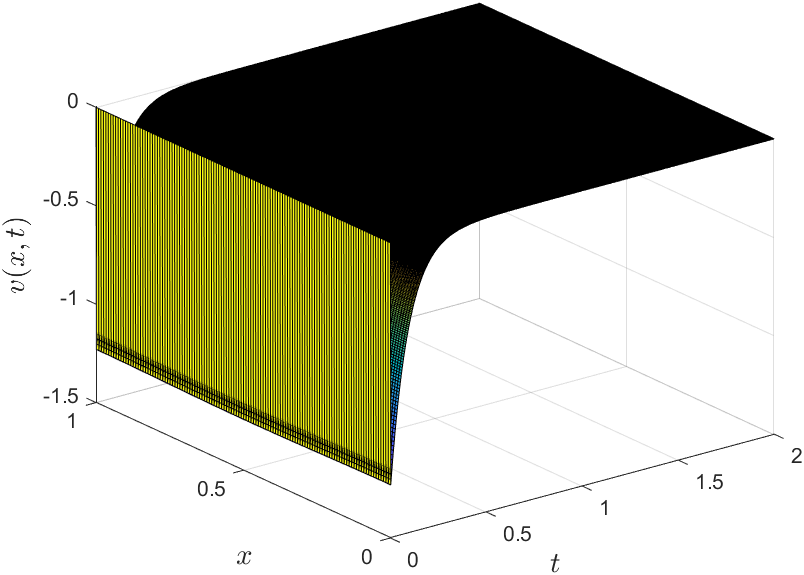}
        \caption{Trajectory of $v(x,t)$ before and after control.}
        \label{fig:v_traj}
    \end{subfigure}
    \caption{Trajectories of the coupled parabolic–elliptic system (1)–(4) 
    with initial condition $u_0 = \sin(\pi x)$.  
    The parameters $\gamma = \tfrac14$, $\rho = \tfrac13$, $\alpha = \tfrac14$, 
    $\beta = \tfrac12$ make the uncontrolled system unstable.  
    Using the control law \eqref{eq:u} with $K=2$, which satisfies 
    the sufficient stability condition \eqref{eq:condition_K}, the closed-loop 
    system becomes stable as predicted by the theory.}
    \label{fig:uv_controlled}
\end{figure}

\begin{figure}[ht!]
    \centering

    \begin{minipage}{0.47\textwidth}
        \centering
        \includegraphics[width=\textwidth]{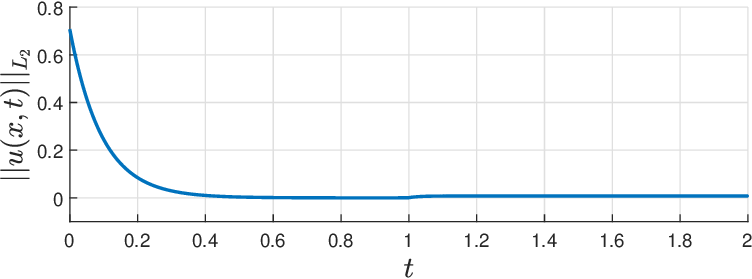}
        \caption*{(a) $\|w(\cdot,t)\|_{L^2}$}
    \end{minipage}
    \hfill
    \begin{minipage}{0.47\textwidth}
        \centering
        \includegraphics[width=\textwidth]{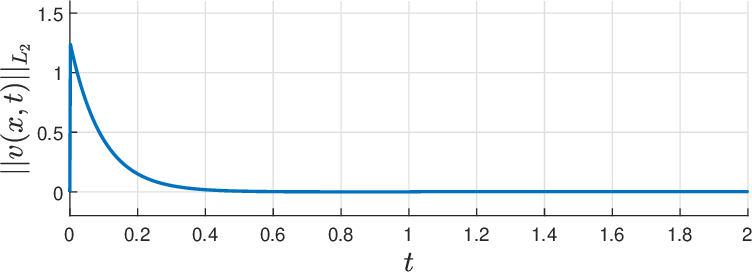}
        \caption*{(b) $\|v(\cdot,t)\|_{L^2}$}
    \end{minipage}

    \caption{The $L^2$-norm of the solutions $w(x,t)$ and $v(x,t)$ for the closed-loop
    system with parameters $\gamma=\tfrac14$, $\rho=\tfrac13$, $\alpha=\tfrac14$,
    $\beta=\tfrac12$, and $c_1 = 1.2$, which satisfies the stability condition (\ref{eq:condition_K}).
    The figure shows the unstable behaviour of the uncontrolled system and demonstrates
    that the control input forces the solutions of the coupled system to decay to zero
    as $t \to \infty$.}
    \label{fig:L2_norms}
\end{figure}

\begin{figure}[ht!]
    \centering

    \begin{minipage}{0.47\textwidth}
        \centering
        \includegraphics[width=\textwidth]{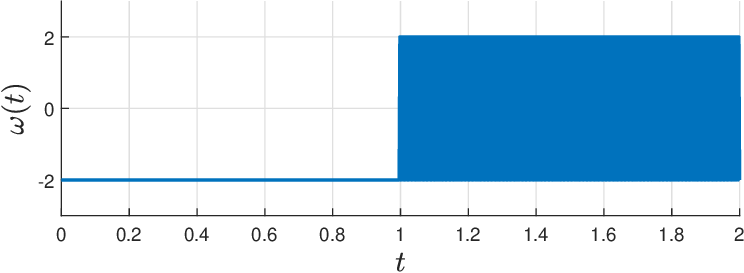}
        \caption*{(a) Control gain signal $\omega(t)$}
    \end{minipage}
    \hfill
    \begin{minipage}{0.47\textwidth}
        \centering
        \includegraphics[width=\textwidth]{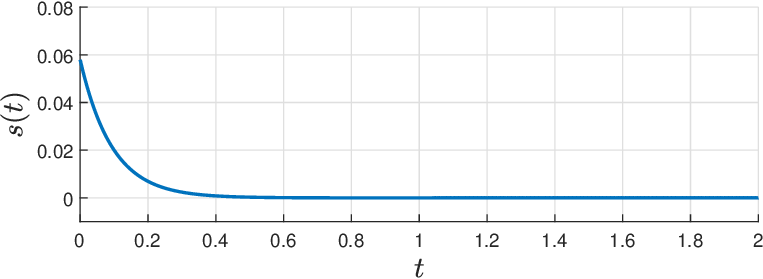}
        \caption*{(b) Sliding mode surface $s(t)$}
    \end{minipage}

    \caption{Control gain signal (\ref{eq:u}) and sliding mode surface for the closed-loop
    system using the parameters $\gamma=\tfrac14$, $\rho=\tfrac13$, $\alpha=\tfrac14$, 
    $\beta=\tfrac12$, and input $\omega$. The plots illustrate the control action 
    and the evolution of the sliding mode surface used in the robust controller design.}
    \label{fig:control_gain_sliding_surface}
\end{figure}

Figure~\ref{fig:uv_controlled} shows the trajectories of the state variables $u(x,t)$ and $v(x,t)$ before and after applying the control law \eqref{eq:u}. Without control, the system exhibits unstable behavior due to the choice of parameters $\gamma=\tfrac14$, $\rho=\tfrac13$, $\alpha=\tfrac14$, and $\beta=\tfrac12$. The introduction of the sliding mode boundary control with gain $K=2$ stabilizes the system in the presence of $d(t) = \sin(20t)$, forcing the trajectories to converge to zero, in agreement with the theoretical predictions. Figure~\ref{fig:L2_norms} depicts the $L^2$-norms of $w(x,t)$ and $v(x,t)$ over time. The plots confirm that the uncontrolled system grows unbounded, while the closed-loop system under the designed control decays to zero, highlighting the finite-time reaching and subsequent stability of the sliding manifold. Figure~\ref{fig:control_gain_sliding_surface} presents the control input $\omega(t)$ and the corresponding sliding mode surface $s(t)$. The control signal exhibits the expected discontinuous behavior to drive the system onto the sliding manifold, while the sliding variable $s(t)$ converges to zero in finite time, demonstrating the robustness of the controller in the presence of bounded perturbations.

\section{Conclusions}
This work addresses the robust stabilization of algebraic parabolic–elliptic systems. We consider matched perturbations, as commonly encountered in boundary state control, which necessitate a robust control strategy. To this end, a sliding mode control  approach is adopted. The stability analysis demonstrates finite-time convergence of the sliding manifold and exponential stability of the closed-loop system. Additionally, due to the discontinuous nature of the closed-loop dynamics, its well-posedness is rigorously established. A numerical example is included to illustrate and validate the effectiveness of the proposed method.

\bibliography{ifacconf}

\end{document}